# Retarding parallel components of a Mueller matrix

IGNACIO SAN JOSÉ[1], JOSÉ J. GIL[2]*

[1] *Instituto Aragonés de Estadística, Gobierno de Aragón, Bernardino Ramazzini 5, 50015 Zaragoza, Spain*
[2] *Department of Applied Physics, University of Zaragoza, Pedro Cerbuna 12, 50009 Zaragoza, Spain*

*Corresponding author: ppgil@unizar.es

**Abstract**

Mueller matrix polarimetry constitutes a nondestructive powerful tool for the analysis of material samples that is used today in an enormous variety of applications. Depolarizing samples exhibit, in general, a complicated physical behavior that requires appropriate mathematical formulation through models involving decomposition theorems in terms of simpler components. In this work, the general conditions for identifying retarding incoherent components of a given Mueller matrix **M** are obtained. It is found that when the coherency matrix **C** associated with **M** has rank **C** = 3,4 it is always possible to identify one or two retarding incoherent components respectively, while in the case where rank **C** =2, such retarding component only can be achieved if and only if the diattenuation and the polarizance of **M** are equal. Since the Mueller matrices associated with retarders have a simple structure, the results obtained open new perspectives for the exploitation of polarimetric techniques in optics, remote sensing and other areas.

## 1. Introduction

Serial and parallel decompositions of depolarizing Mueller matrices have proven to be fruitful frameworks for the study of samples by means of Mueller polarimetry. It is the theory-experiment comparison what leads experimentalists to determine the most appropriate type of decomposition that is more plausible for each situation. In particular, the *arbitrary decomposition* of a Mueller matrix **M** [1-4] provides all possible decompositions of **M** in terms of incoherent parallel combinations of pure (nondepolarizing) Mueller matrices, and the main result of this letter is the determination of the conditions for **M** to contain possible retarding parallel incoherent components as well as the identification of the maximum number of such components.

As a step previous to solving the problem, we introduce the necessary concepts and notations, including the coherency matrix framework, which particularly advantageous for formulating and solving the problem. Let us first recall that a Mueller matrix is said *pure* (or *nondepolarizing*) when it transforms any input totally polarized Stokes vector into a totally polarized Stokes vector. In general, Mueller matrices (pure or not) can be expressed in the block form [5]

$$\mathbf{M} = m_{00}\hat{\mathbf{M}}, \quad \hat{\mathbf{M}} \equiv \begin{pmatrix} 1 & \mathbf{D}^T \\ \mathbf{P} & \mathbf{m} \end{pmatrix},$$

$$\mathbf{m} \equiv \frac{1}{m_{00}} \begin{pmatrix} m_{11} & m_{12} & m_{13} \\ m_{21} & m_{22} & m_{23} \\ m_{31} & m_{32} & m_{33} \end{pmatrix}, \quad (1)$$

$$\mathbf{D} \equiv \frac{(m_{01}, m_{02}, m_{03})^T}{m_{00}}, \quad \mathbf{P} \equiv \frac{(m_{10}, m_{20}, m_{30})^T}{m_{00}},$$

where the superscript $T$ indicates transpose, $m_{00}$ is the *mean intensity coefficient* (i.e. the *transmittance* or *gain* [6-11] of **M** for input unpolarized light), and **D** and **P** are the respective diattenuation and polarizance vectors of **M**. The absolute values of these vectors are the diattenuation $D \equiv |\mathbf{D}|$ and the polarizance $P \equiv |\mathbf{P}|$.

When appropriate, pure Mueller matrices are denoted as $\mathbf{M}_J$. Each $\mathbf{M}_J$ has associated a Jones matrix **T**, a covariance vector **h**, a covariance matrix $\mathbf{H}_J \equiv \mathbf{h} \otimes \mathbf{h}^\dagger$ (where $\otimes$ indicates Kronecker product and the dagger stands for conjugate transpose), a coherency vector **c** and a coherency matrix $\mathbf{C}_J \equiv \mathbf{c} \otimes \mathbf{c}^\dagger$ [12,13]. The relations between the complex elements $t_{ij}$ $(i,j=1,2)$ of **T** and those of **M** are well known and can be found in [13,14]; vector **h** is defined as $\mathbf{h} \equiv (t_1, t_2, t_3, t_4)^T / \sqrt{2}$ where $t_1 \equiv t_{11}$, $t_2 \equiv t_{12}$, $t_3 \equiv t_{21}$, $t_4 \equiv t_{22}$ are the elements of **T**, and vectors **c** and **h** are linked to h by means of $\mathbf{c} \equiv \mathbf{L}\mathbf{h}$ where **L** is the unitary matrix

$$\mathbf{L} \equiv \frac{1}{\sqrt{2}} \begin{pmatrix} 1 & 0 & 0 & 1 \\ 1 & 0 & 0 & -1 \\ 0 & 1 & 1 & 0 \\ 0 & i & -i & 0 \end{pmatrix}. \quad (2)$$

Retarders constitute a particular type of pure Mueller matrices characterized by the following structure

$$\mathbf{M}_R = \begin{pmatrix} 1 & \mathbf{0}^T \\ \mathbf{0} & \mathbf{m}_R \end{pmatrix}, \quad \mathbf{m}_R^T = \mathbf{m}_R^{-1}, \quad \det \mathbf{m}_R = +1, \quad (3)$$

that is, $\mathbf{m}_R$ is a proper orthogonal matrix, and therefore $\mathbf{M}_R$ is a proper orthogonal Mueller matrix. The Jones matrix $\mathbf{T}_R$ associated with $\mathbf{M}_R$ is a unitary matrix, so that the associated coherency vector $\mathbf{c}_R$ has necessarily the form

$$\mathbf{c}_R = (\alpha_1, i\beta_2, i\beta_3, i\beta_4)^T, \quad (4)$$

where $\alpha_1, \beta_2, \beta_3, \beta_4$ are real parameters.

Depolarizing Mueller matrices cannot be associated to Jones matrices, but they always have associated respective covariance and coherency matrices. [13,15]

Given coherency matrix **C**, it always can be expressed as an incoherent sum of pure coherency matrices. Since the coherency matrix is Hermitian, it can be diagonalized as

$$\mathbf{C} = (\operatorname{tr}\mathbf{C})\,\mathbf{U}\operatorname{diag}(\hat{\lambda}_1,\hat{\lambda}_2,\hat{\lambda}_3,\hat{\lambda}_4)\mathbf{U}^\dagger, \qquad (5)$$

where **U** is the unitary matrix whose columns are the eigenvectors of **C**, and $\operatorname{diag}(\hat{\lambda}_1,\hat{\lambda}_2,\hat{\lambda}_3,\hat{\lambda}_4)$ represents the diagonal matrix composed of the ordered nonnegative, intensity-normalized, eigenvalues $\hat{\lambda}_i \equiv \lambda_i/\operatorname{tr}\mathbf{C}$ $(0 \le \hat{\lambda}_4 \le \hat{\lambda}_3 \le \hat{\lambda}_2 \le \hat{\lambda}_1)$. Furthermore, **C** can always be expressed as the following *arbitrary decomposition* in terms of a set of $r$ (where $r \equiv \operatorname{rank}\mathbf{C}$) arbitrary but independent coherency vectors $\mathbf{c}_i$ belonging to the subspace generated by the eigenvectors of **C** with nonzero eigenvalues [1,2]

$$\mathbf{C} = \sum_{i=1}^{r} p_i \mathbf{C}_{Ji}, \quad \mathbf{C}_{Ji} \equiv \mathbf{c}_i \otimes \mathbf{c}_i^\dagger, \quad \operatorname{tr}\mathbf{C}_{Ji} = \operatorname{tr}\mathbf{C},$$

$$p_i = \frac{1}{\sum_{j=1}^{r}\frac{1}{\hat{\lambda}_j}\left|(\mathbf{U}^\dagger \hat{\mathbf{c}}_i)_j\right|^2}, \quad \sum_{i=1}^{r} p_i = 1. \qquad (6)$$

Due to the biunivocal relation between a given Mueller matrix **M** and its associated coherency matrix **C**, the arbitrary decomposition has its Mueller version counterpart

$$\mathbf{M}(\mathbf{C}) = \sum_{i=1}^{r} p_i \mathbf{M}_{Ji}(\mathbf{C}_{Ji}), \quad (\mathbf{M}_{Ji})_{00} = m_{00} = \operatorname{tr}\mathbf{C}, \qquad (7)$$

## 2. Maximun number of retarders as parallel components of a given Mueller matrix

Once the theoretical background has been presented, we are ready to explore the conditions required for **M** to have retarders as components in a decomposition of the form (7). To do so, the equations will be formulated in terms of coherency matrices and we will consider separately the possible cases depending on the value of *r*.

Let us first note that when $r = 1$, then **M** is a pure Mueller matrix, so that its only possible arbitrary component is **M** itself. Furthermore, **M** is a matrix of the form $\mathbf{M}_R$ if and only if **M** lacks of diattenuation and polarizance (let us recall, in passing, that pure Mueller matrices always satisfy $P = D$ [16]).

When $r = 4$, the image subspace of **C**, denoted as range **C**, coincides with all the space of four-component complex vectors and therefore any vector $\mathbf{c}_R$ [se Eq. (4)] can be considered an arbitrary component of **C**. This means that, given a Mueller matrix **M** satisfying $\operatorname{rank}\mathbf{C}(\mathbf{M}) = 4$, then any retarder $\mathbf{M}_{R1}$ (whose associated coherency vector will be denoted as $\mathbf{c}_{R1}$) can be considered an incoherent component of **M**, the corresponding coefficient $p_1$ being calculated by means of Eq. (6), so that

$$\mathbf{M} = p_1 \mathbf{M}_{R1} + (1-p_1)\mathbf{M}_3,$$
$$\mathbf{M}_3 \equiv \sum_{i=2}^{4} p_i \mathbf{M}_{Ji}, \quad \operatorname{rank}\mathbf{C}(\mathbf{M}_3) = 3, \qquad (8)$$

$$p_1 = \frac{1}{\sum_{j=1}^{r}\frac{1}{\hat{\lambda}_j}\left|(\mathbf{U}^\dagger \hat{\mathbf{c}}_{R1})_j\right|^2},$$

where the subscript 3 of $\mathbf{M}_3$ indicates that the coherency matrix $\mathbf{C}_3$ associated with $\mathbf{M}_3$ satisfies $\operatorname{rank}\mathbf{C}_3 = 3$.

The retarding component $\mathbf{M}_{R1}$ can be polarimetrically subtracted from **M** in the following manner [2]

$$\mathbf{M}_3 = \frac{1}{1-p_1}(\mathbf{M} - p_1 \mathbf{M}_{R1}), \qquad (9)$$

where the resulting difference matrix $\mathbf{M}_3$ is a Mueller matrix [2].

Let us now consider a generic Mueller matrix $\mathbf{M}_3$ satisfying $\operatorname{rank}\mathbf{C}(\mathbf{M}_3) = 3$, and denote the associated coherency matrix as $\mathbf{C}_3 \equiv \mathbf{C}(\mathbf{M}_3)$. The image subspace of $\mathbf{C}_3$ (range $\mathbf{C}_3$) is generated by three independent coherency vectors $\mathbf{c}_1,\mathbf{c}_2,\mathbf{c}_3$ (note that a particular set of such independent vectors is that constituted by the eigenvectors of $\mathbf{C}_3$ with nonzero eigenvalue). It is straightforward to prove that, given a set of independent coherency vectors $(\mathbf{c}_1,\mathbf{c}_2,\mathbf{c}_3)$ it is always possible to find (not in a unique manner) three complex coefficients $c_1,c_2,c_3$ such that

$$c_1 \mathbf{c}_1 + c_2 \mathbf{c}_2 + c_2 \mathbf{c}_3 = \mathbf{c}_{R2}, \qquad (10)$$

where $\mathbf{c}_{R2}$ has the specific form shown in Eq. (4) for the coherency vector associated with a retarder, with associated Mueller matrix $\mathbf{M}_{R2}$.

Thus, given $\mathbf{M}_3$ with $\operatorname{rank}\mathbf{C}(\mathbf{M}_3) = 3$, then any retarder $\mathbf{M}_{R2}$ can be considered an incoherent component of $\mathbf{M}_3$, the corresponding coefficient $p_2$ being calculated by means of Eq. (6), so that

$$\mathbf{M}_3 = p_2 \mathbf{M}_{R2} + (1-p_2)\mathbf{M}_2,$$
$$\mathbf{M}_2 \equiv \sum_{i=3}^{4} p_i \mathbf{M}_{Ji}, \quad \operatorname{rank}\mathbf{C}(\mathbf{M}_2) = 2,$$
$$p_2 = \frac{1}{\sum_{j=1}^{r}\frac{1}{\hat{\lambda}_j}\left|(\mathbf{U}^\dagger \hat{\mathbf{c}}_{R2})_j\right|^2}, \qquad (11)$$

where $\hat{\lambda}_j$ are now the eigenvalues of $\mathbf{C}_3/m_{00}$

The pure retarding component $\mathbf{M}_{R2}$ can be polarimetrically subtracted from $\mathbf{M}_3$ in the following manner [2]

$$\mathbf{M}_2 = \frac{1}{1-p_2}(\mathbf{M} - p_2 \mathbf{M}_{R2}). \qquad (12)$$

The only remaining case to be studied is that of Mueller matrices $\mathbf{M}_2$ satisfying $\operatorname{rank}\mathbf{C}_2 = 2$ $[\mathbf{C}_2 \equiv \mathbf{C}(\mathbf{M}_2)]$. In general, the diattenuation $D_2$ and polarizance $P_2$ of $\mathbf{M}_2$ are different and, since matrices $\mathbf{M}_R$ lack of polarizance-diattenuation, it turns out obvious that a decomposition of the form

$$\mathbf{M}_2 = p_3 \mathbf{M}_{R3} + (1-p_3)\mathbf{M}_J, \qquad (13)$$

cannot be realized when $D_2 \ne P_2$, which is not compatible with the fact that a pure Mueller matrix $\mathbf{M}_J$ exhibits equal magnitudes for diattenuation and polarization.

In many theoretical and experimental situations $\mathbf{M}_2$ matrices satisfying $D_2 = P_2 > 0$ appear and therefore it is worth to study such interesting case. As demonstrated in the Appendix, when $\mathrm{rank}\,\mathbf{C}_2 = 2$ and condition $D_2 = P_2 > 0$ is satisfied, then it is always possible to find a Mueller matrix $\mathbf{M}_{R3}$ (associated with a retarder) that can be considered a parallel component of $\mathbf{M}_2$, while the remaining pure diattenuating component $\mathbf{M}_J$ is calculated through the polarimetric subtraction of $\mathbf{M}_{R3}$ from $\mathbf{M}_2$ [2]

$$\mathbf{M}_J = \frac{1}{(1-p_3)}(\mathbf{M}_2 - p_3 \mathbf{M}_{R3}),$$

$$p_3 = \frac{1}{\sum_{j=1}^{r} \frac{1}{\hat{\lambda}_j}\left|\left(\mathbf{U}^\dagger \hat{\mathbf{c}}_{R3}\right)_j\right|^2}, \qquad (14)$$

where $\hat{\lambda}_j$ are now the eigenvalues of $\mathbf{C}_2/m_{00}$.

Obviously, from (14) it follows that the diattenuation and polarizance vectors of $\mathbf{M}_2$ and $\mathbf{M}_J$ are equal (recall that we are considering the case where $D_2 = P_2 > 0$)

$$\mathbf{D}(\mathbf{M}_J) = \mathbf{D}(\mathbf{M}_2), \quad \mathbf{P}(\mathbf{M}_J) = \mathbf{P}(\mathbf{M}_2), \qquad (15)$$

When $D_2 = P_2 = 0$, then the matrix $\mathbf{M}_J$ corresponds to a pure Mueller matrix with zero polarizance-diattenuation that necessarily has the form of a pure retarder. That is to say, when $D_2 = P_2 = 0$, $\mathbf{M}$ can always be decomposed into a convex sum of $r$ pure retarders.

While $\mathbf{D}(\mathbf{M}_J)$ and $\mathbf{P}(\mathbf{M}_J)$ are fully determined, the respective 3×3 submatrices $\mathbf{m}_{R3}$ and $\mathbf{m}_J$ of $\mathbf{M}_{R3}$ and $\mathbf{M}_J$ are determined up to left- and right- respective rotations about the directions of $\mathbf{P}$ and $\mathbf{D}$ respectively [17].

In summary, as indicated in Table I, it has been demonstrated that, given a Mueller matrix $\mathbf{M}$ with $\mathrm{rank}\,\mathbf{C}(\mathbf{M}) = r$ ($\mathbf{C}$ being the coherency matrix associated with $\mathbf{M}$) the maximum number $q$ of retarding incoherent components of $\mathbf{M}$ is

1) $q = 2$ when $r = 4$ and $P \neq D$;
2) $q = 3$ when $r = 4$ and $P = D > 0$;
3) $q = 1$ when $r = 3$ and $P \neq D$;
4) $q = 2$ when $r = 3$ and $P = D > 0$;
5) $q = 0$ when $r = 2$ and $P \neq D$;
6) $q = 1$ when $r = 2$ and $P = D > 0$, and
7) $q = r$ when $P = D = 0$.

Table I. Maximum number $q$ of retarders that can be simultaneous parallel components of a Mueller matrix M with $\mathrm{rank}\,\mathbf{C}(\mathbf{M}) = r$

|           | $r = 4$ | $r = 3$ | $r = 2$ |
|-----------|---------|---------|---------|
| $P \neq D$  | $q = 2$ | $q = 1$ | $q = 0$ |
| $P = D > 0$ | $q = 3$ | $q = 2$ | $q = 1$ |
| $P = D = 0$ | $q = 4$ | $q = 3$ | $q = 2$ |

Thus, from a global point of view, three situations arise,

(a) if $P \neq D$, then the maximum number $q$ of retarders that can be contained in the arbitrary decomposition is $r - 2$ (only applicable when $r \geq 2$), so that any arbitrary decomposition of $\mathbf{M}$ contains at least two diattenuators;

(b) when $P = D > 0$, then $q = r - 1$ and arbitrary decompositions containing an only diattenuator (with diattenuation-polarizance equal that of $\mathbf{M}$) are achievable, and

(c) $P = D = 0$, in which case arbitrary decompositions of $\mathbf{M}$ containing only retarders are possible regardless of the value of $q$.

## Appendix

To simplify further calculations let us apply to $\mathbf{M}_2$ the following *dual-retarder transformation* [18]

$$\mathbf{M}_{2t} = \mathbf{M}_{RO}\mathbf{M}_2\mathbf{M}_{RI},$$

$$\mathbf{M}_{2t} = m_{00}\begin{pmatrix} 1 & D & 0 & 0 \\ D & x_{11} & x_{12} & 0 \\ 0 & x_{21} & x_{22} & x_{23} \\ 0 & 0 & x_{32} & x_{33} \end{pmatrix}, \qquad (16)$$

where $\mathbf{M}_{RI}$ and $\mathbf{M}_{RO}$ are orthogonal Mueller matrices associated with respective *input* and *output* retarders, which can always be determined for any given $\mathbf{M}_2$ and transform it into the *canonical tridiagonal form* $\mathbf{M}_{2t}$. Note that the sings of the transformed elements $x_{01} = x_{10} = D$ are taken positive through the appropriate choice of $\mathbf{M}_{RI}$ and $\mathbf{M}_{RO}$ (the resulting sign of $x_{11}$ being fixed by such choice).

The elements $c_{ij}$ of the coherency matrix $\mathbf{C}_{2t}$ associated with $\mathbf{M}_{2t}$ are given by

$$\begin{aligned}
c_{00} &= \frac{m_{00}}{4}(1 + x_{11} + x_{22} + x_{33}), \\
c_{01} &= c_{10}^* = \frac{m_{00}}{4}\left[2D - i(x_{23} - x_{32})\right], \\
c_{02} &= c_{20}^* = 0, \\
c_{03} &= c_{30}^* = -i\frac{m_{00}}{4}(x_{12} - x_{21}), \\
c_{11} &= \frac{m_{00}}{4}(1 + x_{11} - x_{22} - x_{33}), \\
c_{12} &= c_{21} = x_{12} + x_{21}, \\
c_{13} &= c_{31}^* = 0, \\
c_{22} &= \frac{m_{00}}{4}(1 - x_{11} + x_{22} - x_{33}), \\
c_{23} &= c_{32} = \frac{m_{00}}{4}(x_{23} + x_{32}), \\
c_{33} &= \frac{m_{00}}{4}(1 - x_{11} - x_{22} + x_{33}).
\end{aligned} \qquad (17)$$

Let us now recall that, as it is well known in matrix algebra, given a matrix $\mathbf{N}$ and an arbitrary vector $\mathbf{v}$, then necessarily $\mathbf{Nv} \in \mathrm{range}\,\mathbf{N}$. Therefore given an arbitrary coherency vector $\mathbf{y}$, the vector $\mathbf{z}$ obtained as $\mathbf{z} = \mathbf{C}_{2t}\mathbf{y}$ necessarily belongs to $\mathrm{range}\,\mathbf{C}_{2t}$ and consequently, $\mathbf{z}$ can always be considered as the coherency vector of an incoherent component of $\mathbf{M}_{2t}$, so that we can study if there exists at least one vector $\mathbf{y}$ such that $\mathbf{c}_R = \mathbf{C}_{2t}\mathbf{y}$, where $\mathbf{c}_R$ is a retarding coherency vector [hence with the form shown in Eq. (4)]. This leads to a set of four real equations

$$\mathrm{Im}(c_{R1}) = \mathrm{Re}(c_{R2}) = \mathrm{Re}(c_{R3}) = \mathrm{Re}(c_{R4}) = 0, \qquad (18)$$

where $c_{Ri}$ are the complex components of $\mathbf{c}_R$.

Since $\mathrm{rank}\,\mathbf{C}_{2t} = 2$, the real and imaginary parts of all minors of $\mathbf{C}_{2t}$ are zero. In particular, provided $D > 0$, this condition implies that

$$x_{12}^2 = x_{21}^2,$$
$$(x_{12}+x_{21})(x_{23}+x_{32}) = 0, \quad (x_{12}-x_{21})(x_{23}+x_{32}) = 0, \tag{19}$$

which leads to the three following cases,

a) $x_{12} = x_{21} = 0$,

b) $x_{12} = x_{21}\ (|x_{12}|>0)$, $x_{23} = -x_{32}$,

c) $x_{12} = -x_{21}\ (|x_{12}|>0)$, $x_{23} = -x_{32}$.

For each one of the above cases, vector $\mathbf{z}$ can be expressed as $\mathbf{z} = \mathbf{C}_{2t}\,\mathbf{y}$ in terms of the elements of the respective form of $\mathbf{M}_{2t}$ and of the respective real and imaginary parts $a_j$, $b_j$ of the elements $y_j = a_j + ib_j$ of vector $\mathbf{y}$. This leads to a four equations that involve eight real variables, so that the compatibility of each respective set of four equations can be checked by isolating four variables (taken from $a_j, b_j$ in the most convenient way) and writing them in terms of the four remaining. In particular, it results advantageous to isolate the following variables for each case, a) $a_1, b_2, a_3, a_4$; b) $a_1, b_2, a_2, a_4$, and c) $a_1, b_1, b_2, a_3$. Through this procedure the following respective retarding coherency vectors are obtained, all of them satisfying Eqs. (18)

$$\mathbf{z}_a = \frac{m_{00}}{8D}\begin{pmatrix} a_2\left\{\begin{matrix}4D^2 + (x_{23}-x_{32})^2 \\ -(1+x_{11})^2+(x_{22}+x_{33})^2\end{matrix}\right\} \\[4pt] ib_1\left\{\begin{matrix}4D^2 + (x_{23}-x_{32})^2 \\ -(1+x_{11})^2+(x_{22}+x_{33})^2\end{matrix}\right\} \\[4pt] i2D\left\{\begin{matrix}b_4(x_{23}+x_{32}) \\ +b_3(1-x_{11}+x_{22}-x_{33})\end{matrix}\right\} \\[4pt] i2D\left\{\begin{matrix}b_3(x_{23}+x_{32}) \\ +b_4(1-x_{11}-x_{22}+x_{33})\end{matrix}\right\} \end{pmatrix},$$

$$\mathbf{z}_b = \frac{m_{00}}{8D}\begin{pmatrix} 0 \\[4pt] i\left\{\begin{matrix}-4a_3 x_{12}x_{23}+4b_3 x_{12}D+ \\ b_1\left[4D^2+4x_{23}^2-(1+x_{11})^2+(x_{22}+x_{33})^2\right]\end{matrix}\right\} \\[4pt] i2\left\{\begin{matrix}(a_3 x_{23}-b_3 D)(-1+x_{11}-x_{22}+x_{33}) \\ -b_1 x_{12}(1+x_{11}+x_{22}+x_{33})\end{matrix}\right\} \\[4pt] i2Db_4(1-x_{11}-x_{22}+x_{33}) \end{pmatrix}, \tag{20}$$

$$\mathbf{z}_c = \frac{m_{00}}{8D}\begin{pmatrix} \left\{\begin{matrix}4x_{12}(a_4 x_{23}+b_4 D)+ \\ a_2\left[4D^2+4x_{23}^2-(1+x_{11})^2+(x_{22}+x_{33})^2\right]\end{matrix}\right\} \\[4pt] 0 \\[4pt] i2Db_3(1-x_{11}-x_{22}+x_{33}) \\[4pt] -i2\left\{\begin{matrix}a_2 x_{12}(1+x_{11}-x_{22}-x_{33}) \\ -(a_4 x_{23}+b_4 D)(1-x_{11}-x_{22}+x_{33})\end{matrix}\right\} \end{pmatrix}.$$

Arbitrary values can be assigned to the parameters $a_i$, $b_j$ in each specific case, so that the expressions of vectors $\mathbf{z}_a$, $\mathbf{z}_b$ and $\mathbf{z}_c$ can adopt very simple forms. The general expressions in (20) have been preserved in their general forms for the sake of clarity of the procedure performed.

Therefore, it has been demonstrated that, given $\mathbf{M}_{2t}$, it is always possible to find a coherency vector $\mathbf{c}_R$ with associated Mueller matrix $\mathbf{M}_R$, so that decomposition (13) is realized by replacing $\mathbf{M}_{R3}$ by $\mathbf{M}_{R2}^T \mathbf{M}_R \mathbf{M}_{R1}^T$.